\documentclass[final,5p,times,twocolumn]{elsarticle}
\usepackage{amsmath}
\usepackage{url}
\usepackage{amsthm}

\usepackage{lineno}

\journal{Nucl.~Instrum.~Meth.~A}

\begin{document}

\begin{frontmatter}



\title{A large area detector for thermal neutron flux measurements at the KamLAND site}


\author[IPMU]{A.~Kozlov\corref{cor1}}
\cortext[cor1]{Corresponding author.  Email address: \texttt{alexandre.kozlov@ipmu.jp}}
\author[IPMU]{D.~Chernyak}

\address[IPMU]{Kavli Institute for the Physics and Mathematics of the Universe (WPI), University of Tokyo, Kashiwa, Chiba 277-8583, Japan}
\begin{abstract}
A large area (0.5~m $\times$ 0.5~m) thermal neutron detector was developed based on 0.32~mm-thick $^{6}$LiF:ZnS(Ag) scintillator sheets. The detector was constructed for measurements of seasonal variations in low intensity thermal neutron flux at deep underground sites. The detector response to neutrons was studied using a low intensity $^{241}$Am/$^{9}$Be($\alpha$,n) neutron source. The digital pulse shape discrimination technique was applied to separate neutron capture events from the background caused by traces of $\alpha$-emitters from Uranium and Thorium decay chains in scintillator components. The thermal neutron detection efficiency was 0.368$\pm$0.018 (0.155$\pm$0.009) before (after) event selection based on the digital pulse shape discrimination method. The measured $\alpha$-particle background event rate in the $^{6}$LiF:ZnS(Ag) scintillator was (4.88$\pm$0.06(stat))$\times$10$^{-6}$ $\alpha$ cm$^{-2}$sec$^{-1}$. As a test of detector performance we measured the thermal neutron flux at the KamLAND area of the Kamioka neutrino observatory and compared it with result of the neutron flux measurement carried out at the Super-Kamiokande site. The thermal neutron flux at KamLAND was (6.43$\pm$0.50)$\times$10$^{-6}$ n cm$^{-2}$sec$^{-1}$ that included $\sim$10~\% contribution from neutrons with energies above 1~eV. During the detector construction we identified a source of high amplitude noise pulses generated by 5-inch R1250 photomultipliers manufactured by Hamamatsu Photonics K. K..   
\end{abstract}


\end{frontmatter}

\newpage
\section{Introduction}
\label{sec:intro}
Energetic muons produced in interactions between high energy cosmic rays and atmospheric gases create various types of secondary particles (e.g. fast neutrons \cite{Delorme1995}) or unstable nuclei and deposit energy by ionization in detectors used for rare event search. For that reason, experiments aimed at Dark Matter (DM) detection are typically located deep underground where the cosmic-ray muon flux is highly suppressed. At shallow depths, the muon induced neutron background is dominant compared to other neutron sources: spontaneous fission \cite{Litter1952}, and ($\alpha$,n) reactions on light nuclei (e.g. C, O, F, Na, Mg, Al, Si) caused by traces of $\alpha$-emitters from Uranium (U) and Thorium (Th) decay chains in rock. At certain depths, however, the neutron flux from the ($\alpha$,n) reactions and spontaneous fission becomes dominant. The actual depth where this transition occurs strongly depends on local geology: rock chemical composition and U, Th contents. 

Neutrons are often viewed as a source of critical background for experiments aimed at rare event search. E.g., neutron induced nuclear recoils create background that mimics signal from hypothetical Weakly Interacting Massive Particles - one of the most commonly considered type of DM. It is essential to note that existence of annual modulation in the DM signal caused by the Earth revolution around the Sun \cite{Bernabei2008} is generally viewed as a model independent signature for DM particles. Therefore, factors that may cause seasonal variations in the number of neutron induced background events need to be studied with great care. For example, the water content in rocks that changes periodically (however, with a time lag) following dry and wet seasons, or after melting of snow is one of the factors affecting the neutron energy spectrum and number of neutrons captured inside the rock. The radon activity in the underground air may also have a seasonal component that leads to variations in the number of neutrons produced in ($\alpha$,n) reactions. Effect from these factors may become more complex if underground water contains a high amount of dissolved radon gas or $^{226}$Ra.

The neutron background at deep underground laboratories was investigated by several research groups. The $^{6}$Li-loaded liquid scintillator detectors were used at Modane (France) \cite{Chazal1998} and at Gran Sasso (Italy) underground laboratories \cite{Wulandari2004}. At Baksan Neutrino Observatory (Russia) neutron flux was studied by using a liquid scintillator detector combined with $^{3}$He proportional counters uniformly distributed within the scintillator volume \cite{Abdurashitov2002}. In addition, the thermal neutron flux was measured recently at the Baksan Neutrino Observatory \cite{Alekseenko2017} by using a scintillator detector based on $^6$LiF:ZnS(Ag) components. The $^{3}$He proportional counters were used at Canfranc Underground Laboratory (Spain) \cite{Jordan2013}, Kamioka neutrino observatory (Japan) \cite{Nakahata2005}, and YangYang underground laboratory (South Korea) \cite{Park2013}. 

In the paper, we report detailed information about a new large thermal neutron detector constructed specifically for measurements of seasonal thermal neutron flux variations in a small underground cavity \cite{Kozlov2017} located at the KamLAND area of the Kamioka mine observatory (Hida city, Gifu prefecture, Japan) \cite{Eguchi2003}. The experimental area is about 1000~m underground from the top of the mountain Ikenoyama \cite{Nakahata2005}. The cavity was selected as a location for DM detector consisting of ultra-low background NaI(Tl) crystals \cite{Fushimi2016}.  The Kamioka mine geological structure is rather richly composed of variety of rocks: Limestone, Inishi-rock, skarns, gneiss, ore. As explained in \cite{EnomotoPhDThesis}, the U/Th concentrations in the rock samples collected at the KamLAND area show a wide scatter making it difficult to create a realistic model of neutron production by spontaneous fission and ($\alpha$,n) reactions. Compared with the Japan Island Arc average for U (2.32~ppm), and Th (8.3~ppm) the rock samples at the KamLAND area are depleted in Uranium (0.2-2.6~ppm) and Thorium (0.03-6.0~ppm) giving expectations for a relatively low neutron flux, \cite{EnomotoPhDThesis}. However, existence of a small scale irregularity around a specific location cannot be excluded.  The paper includes a result of the absolute thermal neutron flux measurement performed at the KamLAND site in March, 2018.

\section{The detector design}
\label{sec:design}
Based on the result from the earlier thermal neutron flux measurement carried out in the Kamioka mine (8.26$\pm$0.58)$\times$10$^{-6}$ n cm$^{-2}$sec$^{-1}$ \cite{Nakahata2005}, \cite{Suzuki2009} size of the neutron detector was selected to be relatively large (0.5~m $\times$ 0.5~m). The neutron sensitive part of the detector was made of two 0.32~mm-thick (0.25~m $\times$ 0.5~m) EJ-426HD2-PE2 scintillator sheets produced by Eljen Technology, U.S.A. \cite{Eljen}. The EJ-426HD2-PE2 scintillator is a homogeneous mixture of $^6$LiF and ZnS(Ag) components (mass ratio is 1:2) dispersed in a colorless binder and laminated from both sides by 0.25~mm-thick transparent polyester sheets. To achieve high neutron detection efficiency the Lithium enriched in $^{6}$Li to a 95\% level was used. Due to shortage of $^{3}$He gas supplies neutron detectors based on $^6$LiF and ZnS(Ag) components are recently being tested in various applications as a replacement of $^{3}$He proportional counters (see e.g. \cite{Pino2015}, \cite{Kuzminov2017}). Thermal neutrons are detected via the nuclear reaction

\begin{equation}
 ^{6}Li + n \rightarrow \alpha + t + 4.78\,MeV
\label{eq:reaction}
\end{equation}
\begin{figure}[t]
\centering
\includegraphics[scale=0.1]{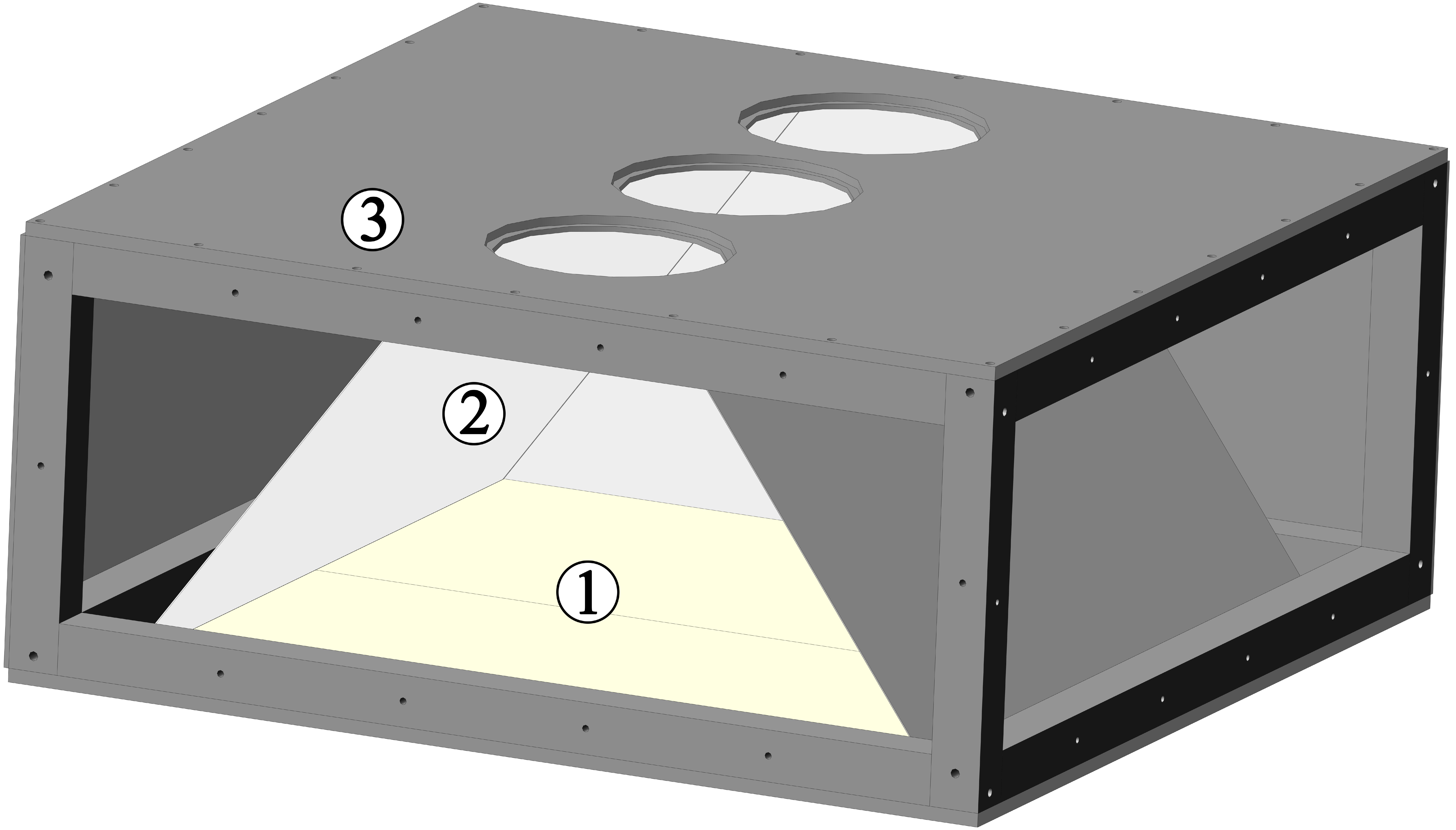}
\caption{The thermal neutron detector structure: 1) two EJ-426HD2-PE2 scintillator sheets; 2) a 0.8~mm-thick aluminum sheets covered by the Tyvek light reflecting paper; 3) an aluminum frame covered with aluminum plates. The inner volume of detector is filled with a pure nitrogen gas.}
\label{fig:Al_frame}
\end{figure}
with a cross-section of 941~barns for 0.025~eV neutrons \cite{ENDF}. Despite of being called the thermal neutron detector it is sensitive to neutrons with higher than thermal energies, e.g., the cross-section for 1~eV neutrons is $\sim$146~barns \cite{ENDF}. A 2.1~MeV $\alpha$-particle and a 2.7~MeV triton produced in the neutron capture reaction interact with the ZnS(Ag) scintillator that emits visible light photons in a wide distribution peaked at 450~nm, \cite{Eljen}. The ZnS(Ag) is a remarkably bright scintillator (95000~photons per MeV, \cite{Eljen}) that produces a signal easily detectable by a detached photomultiplier tube (PMT) positioned at a certain distance from scintillator sheets. We used three 5" Hamamatsu Photonics R1250 PMTs (\cite{HamaR1250}) for detection of the light emitted by the ZnS(Ag) scintillator. The R1250 PMT has a bialkali photocathode with wavelength sensitivity peaked near 400~nm that matches the EJ-426 emission spectrum well.
\begin{figure}[]
\centering
\includegraphics[scale=0.08]{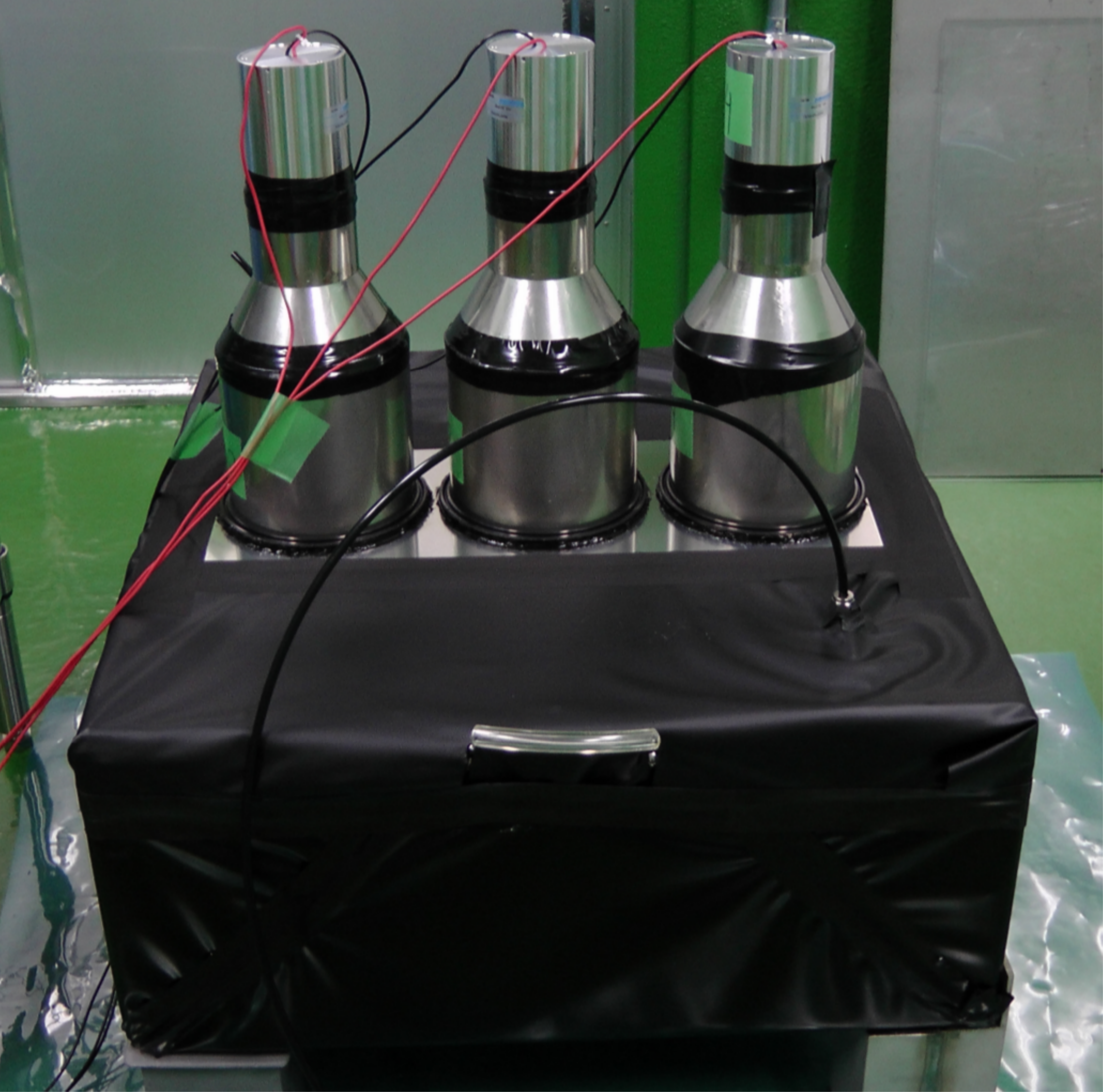}
\caption{A fully assembled thermal neutron detector.}
\label{fig:Neutron_detector}
\end{figure}

Inner structure of the detector is shown in Fig.~\ref{fig:Al_frame}. It is made of an aluminum frame (0.6~m $\times$ 0.6~m $\times$ 0.25~m) covered with aluminum plates. Between the frame and the plates we placed 3-mm-thick PTFE sealant sheets to prevent radon from penetrating inside the detector box. The 8-mm-thick top aluminum plate has three openings sufficient just to fit the R1250 PMT built into the Hamamatsu H6527MOD(K) photomultiplier tube assembly \cite{HamaH6527}. To maximize light collection the bottom aluminum plate was covered by a single Tyvek paper sheet with a 97.9\% light reflection coefficient \cite{Janecek2008} while two EJ-426HD2-PE2 scintillator sheets were attached on top of it. Between the EJ-426HD2-PE2 sheets and the PMTs a light reflector made of a 0.8~mm-thick aluminum sheet also covered by a single layer of Tyvek paper was installed (Fig.~\ref{fig:Al_frame}). The inner volume of detector is filled with a pure nitrogen gas to remove radon. The detector box was wrapped with an anti-static a 0.3~mm-thick black plastic film to avoid light leakage, see Fig.~\ref{fig:Neutron_detector}. 

During test operation of the detector we observed a positive correlation between event rate and radon activity in the room air. Based on the observation, we conclude that presence of the lamination layer on the surface of scintillator did not provide full protection from $\alpha$-particles emitted by the $^{222}$Rn (as well as $^{218}$Po and $^{214}$Po) decay. Perhaps, radon was able to penetrate the scintillator sheets from the edges through microscopic cracks in the mixture of $^6$LiF and ZnS:Ag components. The room where the detector was installed is being continuously supplied with a fresh air taken from the outside of the mine where radon activity is low (typically 15-20~Bq/m$^3$). However, the radon activity in the room air is higher in summer (50-100~Bq/m$^3$) than in winter ($\sim$20~Bq/m$^3$). It happens due to a strong variation in radon activity in the mine air that reaches 3000~Bq/m$^3$ in summer time while concentration in winter is a few tens of Bq/m$^3$ \cite{Nakahata2005}. These variations in the radon activity are caused by seasonal changes in direction of air flow inside the Kamioka mine. Decay of $^{222}$Rn also leads to build up of $^{210}$Po ($\alpha$-emitter, T$_{1/2}$~=~138.4~days \cite{NNDC}) in the scintillator layer that becomes another source of variable in time background. To minimize the background from products of radon decay the neutron detector is being continuously flushed with a 1.5~L/min of pure nitrogen gas. For that purpose, the inner volume of the neutron detector was connected to a centralized source of boiled-off nitrogen gas that supplies nitrogen to the entire KamLAND area, including the KamLAND detector itself. The nitrogen gas flow is being regulated using a mass-flow controller Kofloc 8500MC (\cite{Kofloc8500}) monitored using a data logger Omron ZR-RX45 (\cite{OmronZRRX45}).

\begin{figure}[t]
\centering
\includegraphics[scale=0.8]{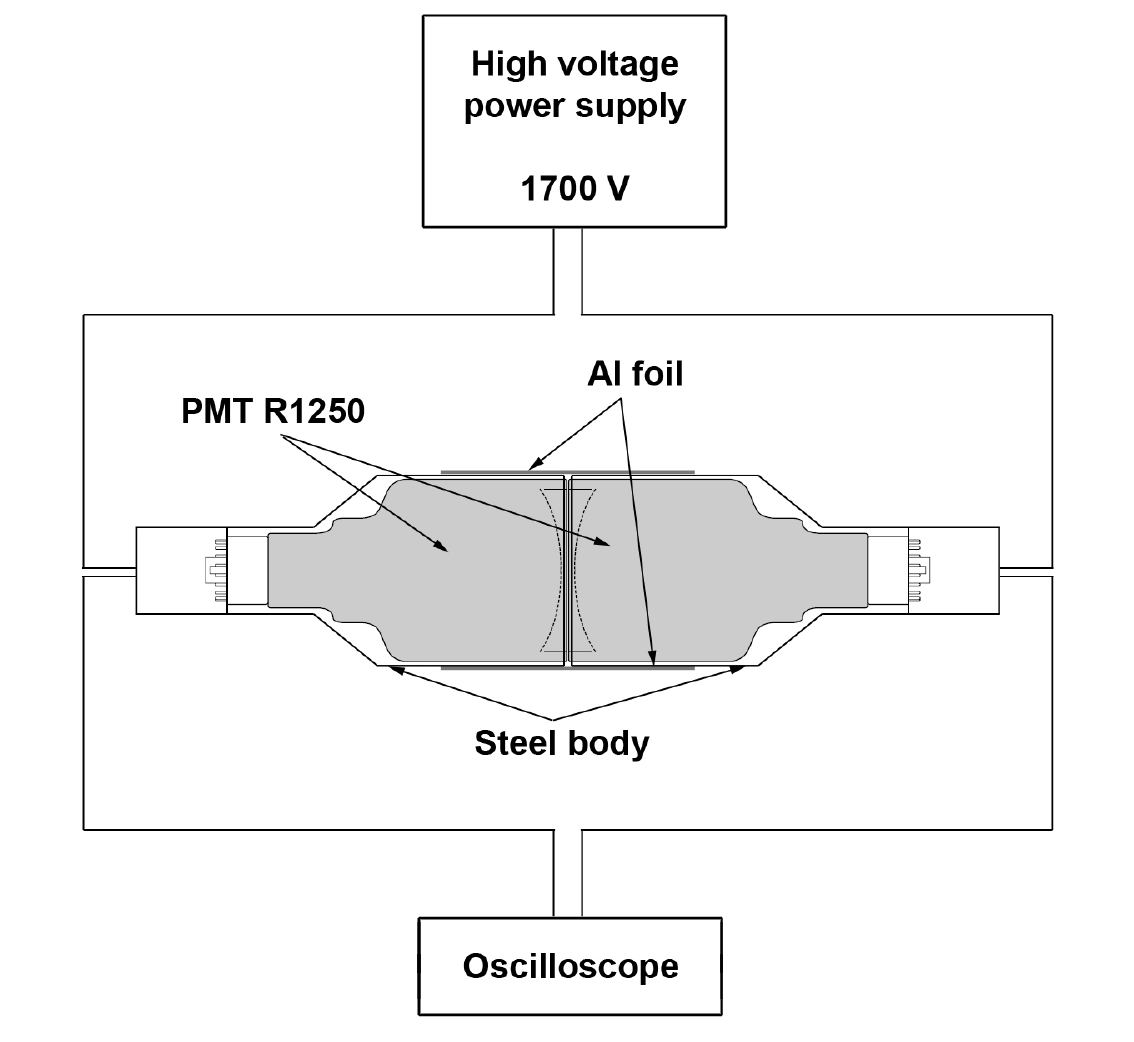}
\caption{The PMT noise test setup: two photomultipliers attached face-to-face with direct optical contact.}
\label{fig:Noise_test}
\end{figure}
\begin{figure}[t]
\centering
\includegraphics[scale=0.3]{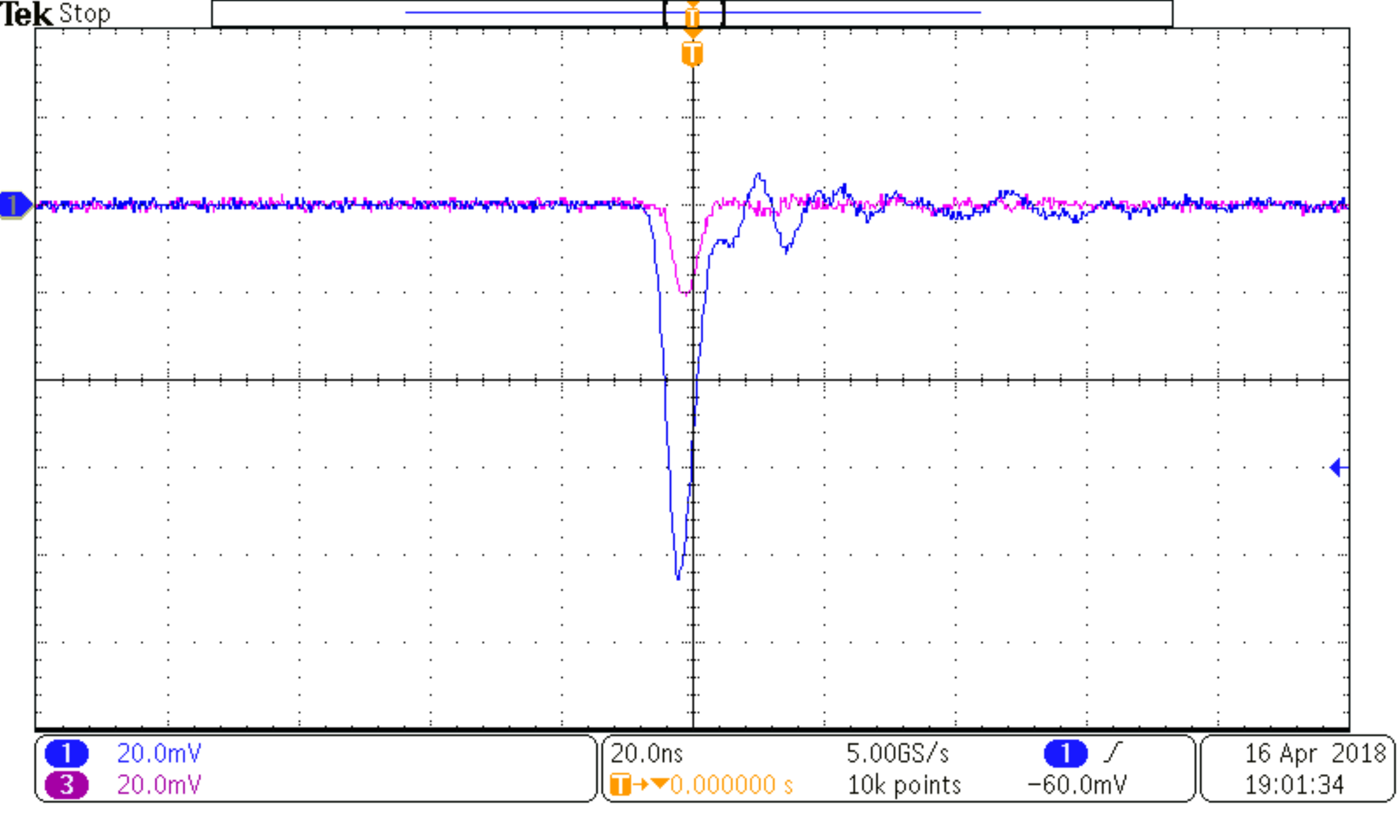}
\caption{Time-correlated high amplitude noise pulses in two optically coupled Hamamatsu Photonics R1250 PMTs connected to CH1 (blue) and CH3 (magenta) of the digital oscilloscope Tektronix MDO3104.}
\label{fig:Noise_pulses}
\end{figure}
The R1250 Hamamatsu PMTs used in the neutron detector were recycled from the liquid scintillator detector called miniLAND that was used to monitor radon activity in the distilled liquid scintillator during the KamLAND Solar neutrino phase, \cite{Konno2008}. Contrary to a standard R1250 PMT these PMTs were produced by Hamamatsu Photonics for use with a positive high voltage that was supplied by a 4-channel CAEN N1470ET NIM-type power supply, \cite{CAENN1470}. The photomultiplier high voltage values were set in a 1484-1520~V range to match gain for three PMTs with a 1~\% accuracy. To tune the high voltage values we performed calibration measurements using a 3-inch NaI(Tl) scintillator crystal attached to the PMTs using an optical grease and a standard $^{137}$Cs calibration source (E$_{\gamma}$~=~661.7~keV). During the measurements PMT with connected NaI(Tl) crystal were placed into a compact 15~cm-thick lead shielding.
 The power supply was monitored and controlled using free CAEN software GECO2020 \cite{GECO2020}. The PMT signals were fed directly into a 4-channel 14-bit 100~MS/s (a 2.25~volt analog input range) CAEN waveform digitizer N6724F (\cite{CAENN6724}).

\section{The photomultiplier noise}
At the beginning of detector development, we used a single 5" Hamamatsu Photonics R1250 PMT for detection of the light emitted by the ZnS(Ag) scintillator. However, we faced a technical problem caused by 8-10~ns wide noise pulses with high amplitudes that composed practically 100\% of data volume. In principle, it was possible to remove the noise during offline data analysis but such solution was far from being optimum due to very fast accumulation of data. Instead, we decided to increase number of PMTs to three and recorded pulses arriving in coincidence only. After the modification a relative fraction of events caused by neutrons and $\alpha$-particles increased to about 15~\% of the data volume. In addition to rejection of the noise, the three phototubes increased efficiency of light collection and minimized signal position dependence. 

For this work we studied noise characteristics of about 10 R1250 photomultipliers. At -1.1~mV threshold and +1500~V high voltage noise rate for best photomultipliers was 70-100~Hz while some PMTs had noise up to 4~kHz. Width of a typical noise pulse was about 8-10~ns. The noise pulse rate and amplitude showed a strong positive correlation with the PMT HV value. Although the manufacturer claimed 3000~V as a maximum supply voltage (\cite{HamaR1250}) practical usage of R1250 PMTs as a photosensor was possible only if HV was limited by about +1500~V (gain $\sim$10$^{6}$) or less. It became a hardware limitation on our ability to adjust a signal amplitude by tuning photomultiplier high voltage.  

In order to understand origin of the noise we measured a cross-talk between noise pulses observed in two neighbor R1250 PMTs. The PMTs were connected to channels 1, and 3 of the 1~GHz digital oscilloscope Tektronix MDO3104 \cite{MDO3104}. During the test, the HV was set to +1700~V. In case of two face-to-face optically coupled R1250 PMTs (see Fig.~\ref{fig:Noise_test}) the second PMT produced a time-correlated noise pulse in 30\% of cases if amplitude of the noise pulse in the channel 1 was higher than 20~mV. Amplitude of noise pulses in the channel 3 was different from pulse-to-pulse but arrival time of two noise pulses agreed within few ns, as shown in Fig.~\ref{fig:Noise_pulses}. Coincidence between noise pulses disappeared after breaking optical contact by placing a black sheet between two PMTs.  

In addition, one of the R1250 PMTs was sent to Hamamatsu Photonics for evaluation. The company personnel performed tests and admitted existence of the noise described here but was not able to explain its origin \cite{Hamamatsu}. 
Earlier we observed a noise with similar characteristics that composed practically 100~\% of data flow for other types of photomultipliers produced by Hamamatsu Photonics such as an ultra-low background 3-inch R11065 \cite{HamaR11065} PMT with a metal body. During the earlier studies the R11065 was placed into an ultra-low background passive shielding made of a 15~cm-thick lead and 5~cm-thick pure copper layers flushed with a pure nitrogen gas and located deep underground at the Kamioka mine. Based on these observations we conclude that this type of noise was caused by fast flashes of light emitted inside the PMT itself. We think that only a fraction of the noise can be explained by the Cherenkov light emitted by energetic electrons produced by environmental $\gamma$-rays in the PMT materials. 

Despite of a weak optical coupling between the neighbor PMTs placed vertically on top of the neutron detector box the coincidence rate for noise pulses in two PMTs was still high, and we decided to use a triple coincidence online trigger to resolve the problem. More detailed discussion about the cause of light emission by Hamamatsu Photonics phototubes is out of scope of this work and will be reported separately elsewhere.   

\section{The data taking and analysis}
\label{sec:data_taking}
\begin{figure}[t]
\centering
\includegraphics[scale=0.42]{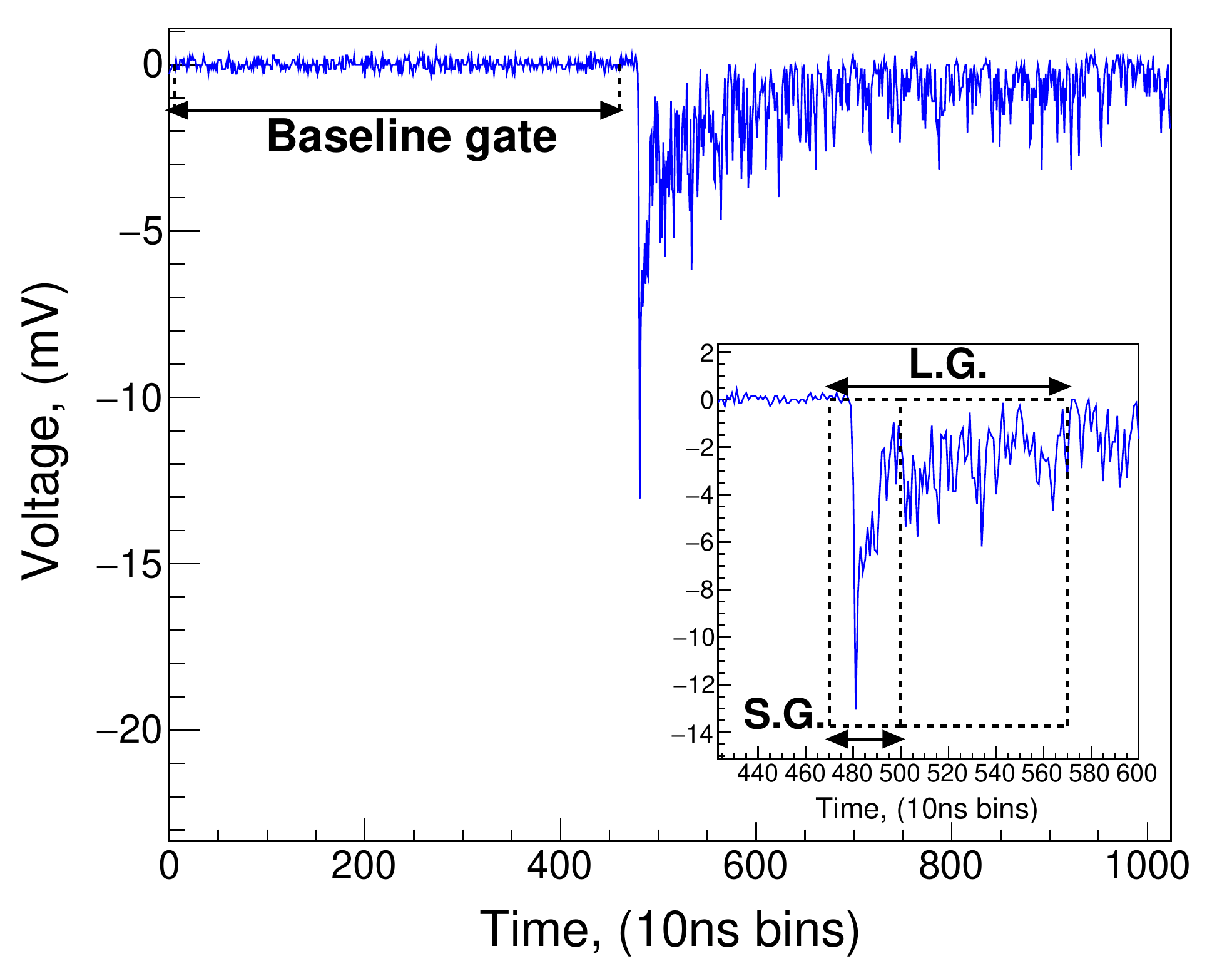}
\caption{A typical PMT signal for the neutron capture event: a fast scintillation component is integrated over a Short Gate (S.G.) which is a 0.3~$\mu$sec long; the Long Gate (L.G.) is a 1~$\mu$sec long; an averaged baseline value is calculated over the 4.6~$\mu$sec baseline gate.}
\label{fig:Waveform}
\end{figure}

The data recording was done using a free CAEN software application called WaveDump \cite{CAENWDump}. The WaveDump online software trigger was set to record signals from three phototubes that: a) exceeded individual thresholds (-1.5~mV from the mean of baseline offset distribution); b) arrived in coincidence within a 100~nsec time window. Data were first stored in the ASCII format, and converted into a ROOT-tree format (\cite{CERNROOT}) offline. The PMT signal detection window was a 10~$\mu$sec long. The first 4.6~$\mu$secs prior a trigger (trigger starts at the center of the detection window) were used to calculate a baseline offset value event-by-event and channel-by-channel. A typical PMT signal (baseline subtracted) from a neutron capture event is shown in Fig.~\ref{fig:Waveform}. 

A thin layer of $^{6}$LiF:ZnS(Ag) scintillator is practically transparent to environmental $\gamma$-rays. Moreover, energetic electrons produced by $\gamma$-ray scattering or by $\beta$-emitters deposit only a small fraction of their energy in the scintillator. Therefore, the light yield emitted by ZnS(Ag) for energetic electrons is low. Contrary to electrons, $\alpha$-particles deposit energy by ionization in a very short range within the scintillator. For that reason $\alpha$-particles become a source of bright signals that overlap with signals from the neutron capture. We applied Pulse-Shape Discrimination (PSD) method to achieve the best possible separation between signals from the neutron capture and the $\alpha$-particle background. The PSD parameter was computed event-by-event as 
\begin{equation}
PSD = \frac{\Sigma \,(L.G.) - \Sigma \,(S.G.)}{\Sigma \,(L.G.)}
\label{eq:psd}
\end{equation}
where $\Sigma \,(L.G.)$ is the sum of signals from three PMTs over a 1~$\mu$sec Long Gate, $\Sigma \,(S.G.)$ is the sum of signals over a Short Gate that is a 0.3~$\mu$sec long, as demonstrated in Fig.~\ref{fig:Waveform}. Duration of gate time windows was chosen by minimizing overlap between distributions of almost pure samples of neutron capture events originated from the $^{241}$Am/$^{9}$Be($\alpha$, n) source and $\alpha$-particle induced background events.

\begin{figure}[t]
\centering
\includegraphics[scale=0.42]{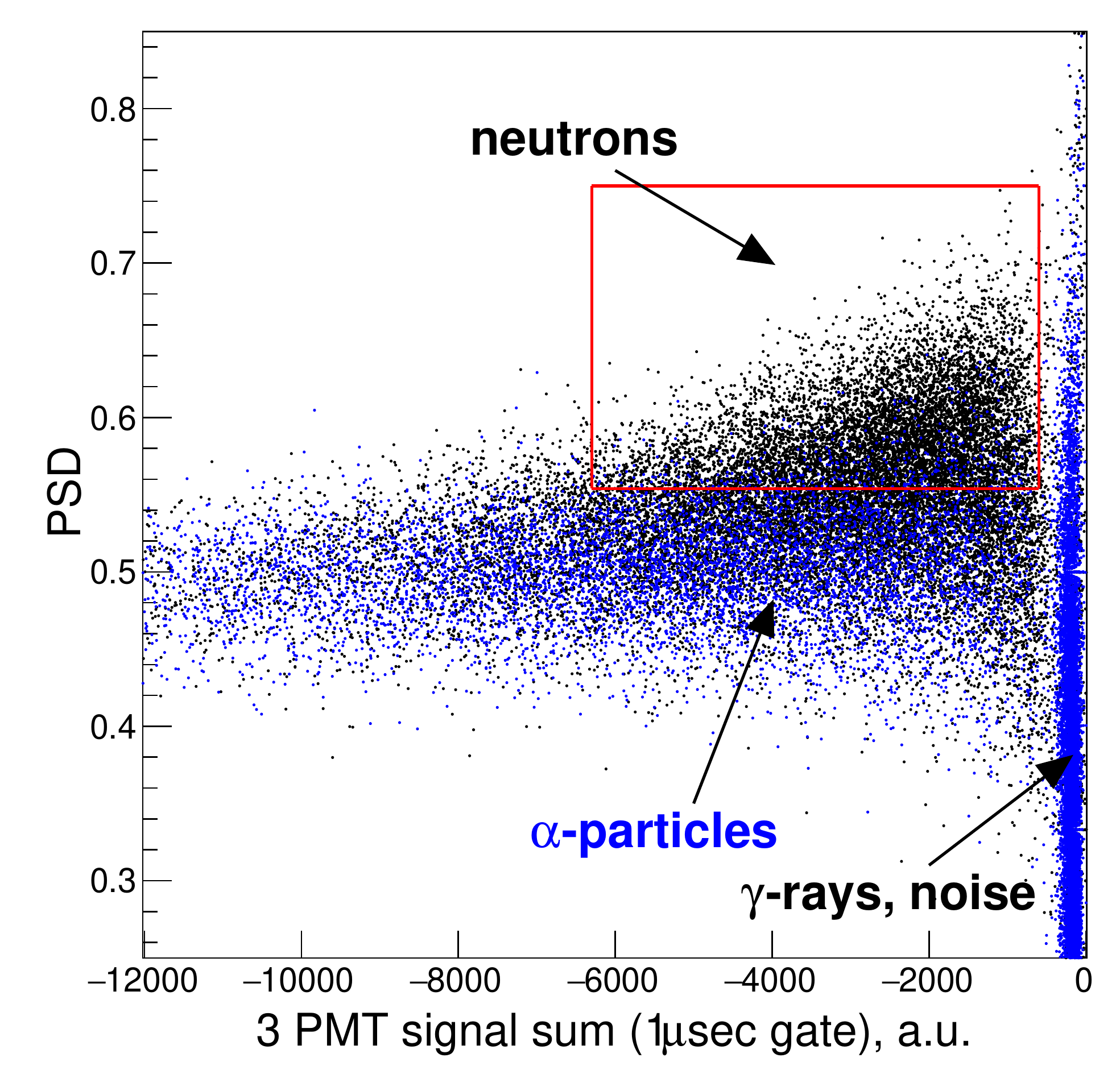}
\caption{The $^{241}$Am/$^{9}$Be neutron source calibration data (black histogram), the $\alpha$-particle background (blue histogram); a red box shows boundaries of software cuts applied to select neutron capture events.}
\label{fig:neutron_alpha}
\end{figure}
The $\alpha$-particle induced background was measured using an external neutron shielding assembled around the detector out of polyethylene bricks (50~mm $\times$ 100~mm $\times$ 200~mm) loaded with Boron (20\%). To avoid overestimation of the $\alpha$-particle background we used the GEANT4 simulation package \cite{GEANT4} for calculation of the thermal neutron flux created by moderation of fast neutrons emitted from rocks after passing through 5 and 15~cm-thick layers of Boron loaded polyethylene. In the GEANT4 model the fast neutron energy was set to 5~MeV that is higher than the maximum of the fast neutron energy distribution (located at 3~MeV) measured at Modane \cite{Chazal1998}. From the simulations we found that the fraction of fast neutrons that had kinetic energy shifted below 1~eV and not captured by the 5 and 15~cm-thick neutron shielding was 3.2$\times$10$^{-4}$ and 2.8$\times$10$^{-4}$, respectively. Based on the simulation result and the fast neutron flux value measured at the Kamioka mine (1.2$\times$10$^{-5}$ n cm$^{-2}$sec$^{-1}$ by \cite{Nakahata2005}) we determined that a 10-15~cm-thick neutron shielding made of the Boron loaded polyethylene effectively cuts off all environmental thermal neutrons while number of fast neutrons moderated to thermal energies and not captured by the shielding is negligibly small. The measured background rate due to traces of $\alpha$-emitters in the scintillator sheets, R$_{\alpha}$, was 12.2$\pm$0.1(stat)~mBq (condition $\Sigma \,(L.G.)$~$<$~-600 for noise rejection was applied).

The detector response to neutrons was studied using the $^{241}$Am/$^{9}$Be($\alpha$, n) calibration source positioned under the bottom of detector box in a moderator made of plastic container filled with a pure water. The $^{137}$Cs standard calibration source (activity is $\sim$250~kBq, E$_{\gamma}$~=~661.7~keV) was used to understand sensitivity of the detector to $\gamma$-rays. During calibration measurements the bare $^{137}$Cs source was positioned under the bottom of detector box. 

\begin{figure}[t]
\centering
\includegraphics[scale=0.44]{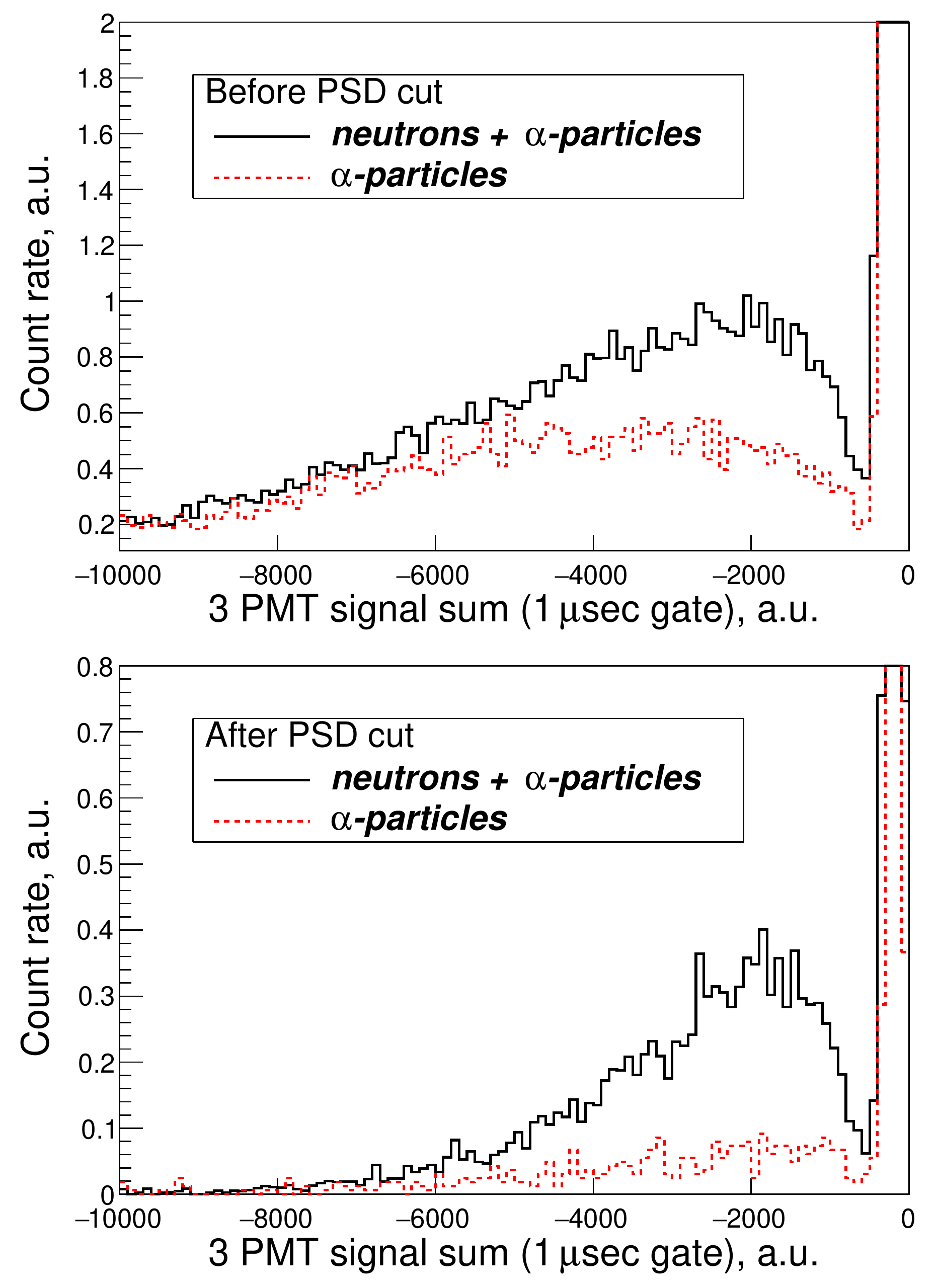}
\caption{(Top) A raw data sample measured at the Kamioka mine; (Bottom) the same data sample after the 0.554~$<$~PSD~$<$~0.75 selection cut used to suppress the $\alpha$-particle background.}
\label{fig:PSD_cut}
\end{figure}

The Fig.~\ref{fig:neutron_alpha} shows values of the PSD parameter versus the $\Sigma \,(L.G.)$ for events measured with the $^{241}$Am/$^{9}$Be neutron source and the $\alpha$-particle background events (black and blue scatter histograms, respectively). Signals from the neutron capture, in general, are characterized by higher PSD values and lower amplitudes compared to signals from $\alpha$-particles. The Fig.~\ref{fig:neutron_alpha} also demonstrates that events caused by the PMT noise, electronics noise and environmental $\gamma$-rays can be rejected by setting a certain threshold value on $\Sigma \,(L.G.)$~$<$~-600. We optimized expression S/$\sqrt(B)$ to determine best possible PSD selection values, where $S$ is the number of neutron capture events originated from rocks around the cavity at the Kamioka mine (after the $\alpha$-background subtraction) and $B$ is the number of $\alpha$-particle background events. The PSD cut effectiveness to suppress the $\alpha$-induced background is demonstrated in Fig.~\ref{fig:PSD_cut} that shows a data sample acquired at the Kamioka mine without and with the PSD cut selection: 0.554~$<$~PSD~$<$~0.75. The best separation between neutrons and the background composed of $\alpha$-particles, noise and $\gamma$-rays was achieved for events inside the box cut shown in Fig.~\ref{fig:neutron_alpha}. After the box cut event rate caused by $\alpha$-emitters in the scintillator sheets, R$_{\alpha}$(s), was only 0.679$\pm$0.034(stat)~mBq (reduction by a factor~$\sim$18) while the box cut efficiency for neutron capture events remained relatively high ($\sim$42\%). 

\section{Tests of the detector performance}
\label{sec:neutron_flux}

\begin{figure}[t]
\centering
\includegraphics[scale=0.44]{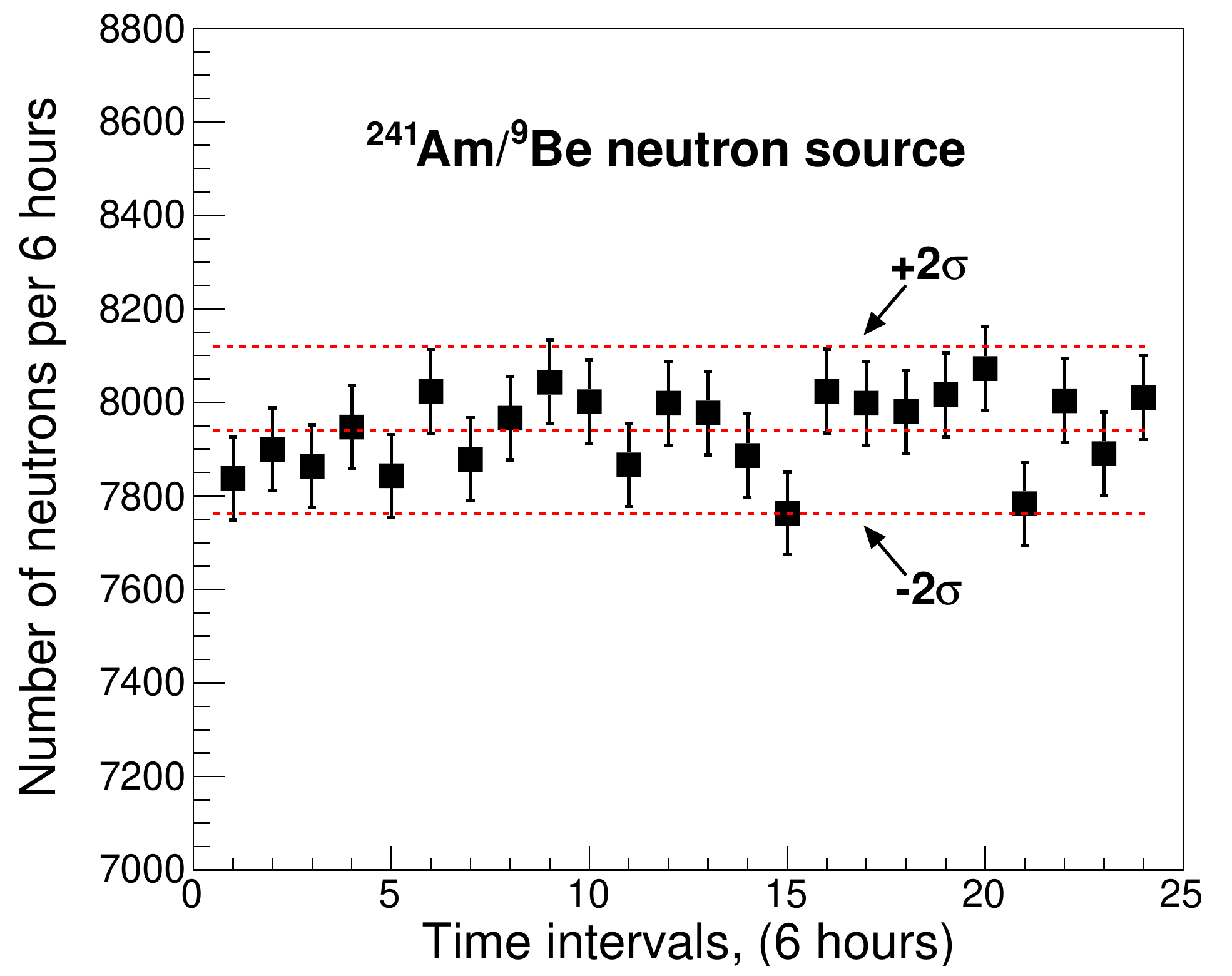}
\caption{The measured neutron event rate during a 6~day long stability test.  A central dashed line shows the mean neutron rate value during the entire 6~day period while two other dashed lines show the expected $\pm 2\sigma$ range due the statistical uncertainty.}
\label{fig:stability_test}
\end{figure}

Working characteristics of the neutron detector were determined by the following measurements: a) a long-term neutron detection stability test, and b) the thermal neutron flux measurement at the Kamioka mine. It is essential to note that the neutron detector operates in steady environmental conditions at the underground laboratory. An air conditioning system in the room maintains temperature at 18.0$\pm0.5$~deg Celsius that keeps parameters of DAQ electronics stable and prevents variations of the photomultipliers gain. To demonstrate an overall stability of the neutron detector we performed a 6~day long measurement using the $^{241}$Am/$^{9}$Be($\alpha$, n) calibration source positioned under the bottom of detector box inside the water moderator. Result of the test is demonstrated in Fig.~\ref{fig:stability_test} where each data point corresponds to the number of neutrons detected during a 6~hour time period (the software box cut shown in Fig.~\ref{fig:neutron_alpha} was applied). The central horizontal dashed line (see Fig.~\ref{fig:stability_test}) shows the mean neutron detection rate averaged over the entire 6~day period while two other dashed lines show expected the $\pm 2\sigma$ statistical uncertainty region of the mean value. The stability test demonstrated that variations in the number of detected neutrons observed during the test can be explained by statistical fluctuations only. Based on the test result we concluded that overall systematic factors such as instability of electronics and PMT gain variations are insignificant compared with the statistical uncertainty and can be neglected.
         
Measurement of the absolute thermal neutron flux requires knowledge of the absolute neutron detection efficiency that can be either calculated or measured experimentally. As a solution, we considered two options: calculation of the detection efficiency using a realistic model of the detector and a direct measurement using the $^{241}$Am/$^{9}$Be calibration source. However, in order to create a realistic GEANT4 model that includes photon transport from a scintillation point to the photomultiplier tubes one needs to know a true distribution of dimensions for $^6$LiF and ZnS(Ag) particles in the scintillator sheets. According to the manufacturer a typical particle size for the ZnS(Ag) powder is 8~$\mu$m \cite{ZnS:Ag}, while the $^6$LiF particle size is unknown. However, a magnified image of almost identical product (a 0.5~mm-thick EJ-426HD2 sheet with $^6$LiF:ZnS(Ag), mass ratio 1:2 \cite{Ely2013}) shows that particles for both $^6$LiF and ZnS(Ag) have size between 1 and 10~$\mu$m and smaller particles are likely to exist too. According to \cite{Yehuda-Zada2014} the 2.1~MeV $\alpha$-particle has a very short stopping range (6-7~$\mu$m) in $^6$LiF or ZnS(Ag) and $\sim$9~$\mu$m in the binder while 2.7~MeV tritons have much longer stopping range: 32-33~$\mu$m in $^6$LiF or ZnS(Ag) and 51-52~$\mu$m in the binder. Without knowing the true size distribution for the $^6$LiF:ZnS(Ag) particles it is difficult to calculate correctly 
\begin{itemize}
\item a fraction of energy that tritons and $\alpha$-particles deposit inside the ZnS(Ag) scintillator grains, and thus number of photons emitted along a particle track;
\item light attenuation in the ZnS(Ag) particles and, therefore, expected light yield and spectrum of photons exiting the scintillator layer and determine fraction of events that are too dim to be detected by the PMTs with the selected signal threshold. 
\end{itemize}
Due to these uncertainties we decided not to use this method. 

A direct measurement of the neutron detection efficiency requires a calibrated neutron source with a well known intensity and neutron energy distribution. We determined these properties of the $^{241}$Am/$^{9}$Be neutron source using a combination of the experimental data taken during deployment of the neutron source into the KamLAND neutrino detector and the GEANT4 modelling. Due to a very large mass (1000~tons of a liquid scintillator consisting of mineral oil (80\%) and pseudocumene (20\%) \cite{Eguchi2003}) practically all neutrons emitted by the $^{241}$Am/$^{9}$Be source inside KamLAND were captured and produced a signal. In the liquid scintillator 99.46\% of neutrons were captured on protons via the nuclear reaction n + p $\rightarrow$ $^2$H + $\gamma$ (2.2~MeV). Detection of signals from the 2.2~MeV $\gamma$-rays allowed to measure the absolute neutron flux emitted by the source regardless of the neutron energy. The neutron source was sealed in a 5~mm-thick, a 18~mm-in-diameter acrylic disk. During deployments into the KamLAND the $^{241}$Am/$^{9}$Be neutron source was placed into a small stainless steel container shaped as a cylinder: a 38~mm in diameter and 22~mm high. Between the acrylic disk and walls of the container a 2.25~mm-thick layer of lead shielding was laid. The stainless steel container was surrounded by a cylindrical moderator made of a polyethylene (a 104~mm in diameter and 88~mm high). The neutron source activity, $I[Bq]$, was calculated as 
\begin{equation}
I[Bq] = \frac{ N_{\gamma} }{t \cdot f_{abs}} \cdot f_{\delta t} 
\label{eq:AmBe_intensity}
\end{equation}
where $N_{\gamma}$ is the number of 2.2~MeV $\gamma$-rays detected during source deployment at KamLAND (after subtraction of accidentals and background from proton recoils), t - is time of the measurement in seconds, f$_{abs}$ is a fraction of neutrons that produced a signal in the KamLAND scintillator within the energy range used for selection of 2.2~MeV $\gamma$-rays in the experimental data, and f$_{\delta t}$ - is the correction factor that takes into account reduction in the $^{241}$Am (T$_{1/2}$~=~432.6~years, \cite{NNDC}) activity since time of the measurement (15~years). The f$_{abs}$ correction was calculated using the GEANT4 model that includes the neutron source, the polyethylene moderator and the KamLAND detector. Use of the KamLAND data allowed to determine the absolute neutron activity of the source with a good accuracy $I$~=~74.6 $\pm 0.3$(stat) $\pm 1.9$(syst)~Bq, where the systematic error was estimated using different models of the background under the peak produced by 2.2~MeV $\gamma$-rays. 

The energy distribution of neutrons emitted by the source was more uncertain. In the $^{241}$Am/$^{9}$Be source neutrons are produced predominantly via nuclear reactions $\alpha$ + $^{9}$Be $\rightarrow$ $^{12}$C$^*$ + n, where $^{12}$C nucleus can be in the ground or excited state.  In the earlier work \cite{Araki2006} the neutron energy spectra measured at KamLAND were compared to the GEANT simulation result obtained using the neutron energy spectra from \cite{Vijaya1973} and good agreement was found for transitions to the ground, 1$^{st}$ and 2$^{nd}$ excited states of $^{12}$C. However, the scintillation light yield for recoiled protons and $^{12}$C nuclei produced by fast neutron scattering in organic scintillators is strongly suppressed by the effect called quenching \cite{Birks1951}. Due to quenching, signals from a low energy part of the neutron spectrum originated from a three-body breakup reaction $^{9}$Be($\alpha$, $\alpha'$) $\rightarrow$ $\alpha$ + $\alpha$ + n were not observed at KamLAND and remained unverified. In addition, the neutron spectrum published in \cite{Vijaya1973} gives no information about neutrons with energy below 0.1~MeV that is likely to cause some underestimation of the expected thermal neutron flux value. From the GEANT4 simulations we found that despite of a relatively small contribution ($\sim$7.3~\%) to the neutron energy distribution given in \cite{Vijaya1973} the three-body breakup reaction gives $\sim$20~\% of all thermal neutrons after passing through the water moderator. We conservatively assigned a $\pm$20~\% systematic uncertainty to the neutron flux originated from a three-body breakup reaction that resulted in a $\pm$4~\% systematic error of the calculated thermal neutron flux value.

During the neutron detection efficiency measurements the $^{241}$Am/$^{9}$Be neutron source was positioned behind the moderator made of a 1~L plastic container filled with a water. Use of the water moderator allowed to produce a sufficiently high thermal neutron flux. At first, the detector was positioned about 25~cm above the room's floor made of a 10~cm-thick concrete layer on top of the rock. The moderator and the neutron source were placed on the floor under the detector. The GEANT4 model was prepared that describes the neutron detector, the moderator with the neutron source and the room itself. However, we found that uncertainties in the chemical composition and water content of the concrete floor and surrounding rocks create a large systematic error in the thermal neutron flux passing through the detector. Therefore, during the neutron detection efficiency measurements we placed the detector into a 10-15~cm-thick shielding made of Boron loaded polyethylene to isolate it from the surrounding environment and minimize systematic uncertainties in the GEANT4 simulations. The moderator and the neutron source were placed onto the 15~cm-thick bottom layer of Boron loaded polyethylene. To measure position dependence in the neutron detection efficiency 9 positions of the neutron source were used: under the detector center, and shifted by $\pm$15~cm along both axes of the detector. 

By using the GEANT4 modelling we determined a fraction of neutrons ($f_{th}$) emitted by the $^{241}$Am/$^{9}$Be source that: a) passed through the layer of the $^{6}$LiF:ZnS(Ag) scintillator; b) was moderated to thermal energies (below 1~eV). E.g., the $f_{th}$ was 1.49~\% for the neutron source located under the detector center. To take into account contribution from neutrons with energies above 1~eV the $f_{th}$ was multiplied by a correction factor calculated from the simulated neutron energy distribution weighted by the probability of neutron capture on $^{6}$Li. After the correction the $f_{th}$ value increased to a 1.65~\% for the neutron source positioned under the detector center.

Based on the $^{241}$Am/$^{9}$Be calibration data and the GEANT4 simulation result we calculated the absolute thermal neutron detection efficiency ($\epsilon$), as
\begin{equation}
\begin{split}
\epsilon^{i} = \frac{ R_{AmBe}^{i} - R_{\alpha}}{R_{th}^{i}} \\
\epsilon = \frac{1}{9}\sum_{i=0}^{8} {\epsilon^{i}} 
\label{eq:det_eff}
\end{split}
\end{equation}
where $\epsilon^{i}$ is the thermal neutron detection efficiency for 9 locations of the neutron source identified by the $i$~=~0,1, ..,8 symbols, $R_{AmBe}^{i}$ is the number of events from the $^{241}$Am/$^{9}$Be source per second remained after the software cut ($\,\Sigma \,(L.G.)<$-600) used to reject noise and $\gamma$-ray events, $R_{\alpha}$ is the number of events per second caused by environmental $\alpha$-particles (introduced in Section~\ref{sec:data_taking}), $R_{th}^{i}$ is the expected number of thermal neutrons per second crossing the $^{6}$LiF:ZnS(Ag) scintillator layer. The $R_{th}^{i}$~=~$I \cdot f_{th}^{i}$ was calculated using the absolute activity $I$ of the neutron source and the $f_{th}^{i}$ value at each position of the neutron source. The thermal neutron detection efficiency, $\epsilon$, was equal to 0.368$\pm$0.018, where uncertainty included: 1) a combined uncertainty, $\delta_{c}$, due to the statistical uncertainty in the number of detected $^{241}$Am/$^{9}$Be events, and uncertainty due to position dependence in the neutron detection efficiency; 2) the full uncertainty in the absolute activity of the neutron source, $\delta_{I}$; and 3) the uncertainty in the simulated $f_{th}$ value, $\delta_{fth}$ .
We defined $\delta_{c}$ as $\max |{\epsilon - \epsilon^{i}}|$ , $i$=0,1, ..,8. A small value of $\delta_{c}$~=~1.39\% proved that $\epsilon$ was practically uniform over the entire area of the detector. The $\delta_{I}$~=~2.59\% was calculated as the quadrature of the statistical and systematic errors in the absolute activity of the neutron source $I$. The $\delta_{fth}$~=~4.0\% was due to the systematic uncertainty in the thermal neutron flux originated from the three-body breakup reaction $^{9}$Be($\alpha$, $\alpha'$) $\rightarrow$ $\alpha$ + $\alpha$ + n, as was explained earlier in the text. From the Eq.~\ref{eq:det_eff} one can see that $\epsilon$ is a mean probability for a thermal neutron passed through the scintillator layer to produce a signal exceeding the software cut ($\,\Sigma \,(L.G.)<$-600) threshold. It is a general characteristic of the detector that does not depend on particular software selection criteria.

To achieve a high S/N ratio in presence of the background caused by $\alpha$-particles and low underground neutron flux we applied the software box cut shown in Fig.~\ref{fig:neutron_alpha}. A total thermal neutron detection efficiency, $\epsilon_{t}$, that included the software box cut efficiency was calculated using the same calibration data taken at 9 positions of the $^{241}$Am/$^{9}$Be neutron source: 
\begin{equation}
\begin{split}
\epsilon_{t}^{i} = \frac{ R_{AmBe}(s)^{i} - R_{\alpha}(s)}{R_{th}^{i}} \\
\epsilon_{t} = \frac{1}{9}\sum_{i=0}^{8} {\epsilon_{t}^{i}} 
\label{eq:det_eff_soft}
\end{split}
\end{equation}
where $\epsilon_{t}^{i}$ is the total thermal neutron detection efficiency calculated for 9 locations of the neutron source ($i$=0,1, ..,8), and $R_{AmBe}(s)^{i}$ is the number of events from the $^{241}$Am/$^{9}$Be source per second remained after the software box cut. The total thermal neutron detection efficiency, $\epsilon_{t}$, was equal to 0.155$\pm$0.009, where uncertainty was calculated in the same way as uncertainty in $\epsilon$, except that $\delta_{c}$~=~$\max |{\epsilon_{t} - \epsilon_{t}^{i}}|$ , $i$~=~0,1, ..,8. The value of $\delta_{c}$ for $\epsilon_{t}$ was 2.89\%, that is higher compared to the same uncertainty for $\epsilon$ but still low compared to the systematic uncertainty in $R_{th}^{i}$.

In addition, we estimated the following factors capable of reducing the thermal neutron flux passing through the scintillator layer: 
\begin{itemize}
\item neutron captures in the window and body of R1250 photomultiplier made of borosilicate glass containing Boron oxide (the $^{10}$B neutron capture cross-section for 0.025~eV neutrons is 3851~barns \cite{ENDF}, while $^{10}$B natural abundance is $\sim$20\%); 
\item thermal neutrons scattering off and capture in materials of the detector box made of Aluminum alloys 5052 and 6063 \cite{Aluminum};
\item the thermal neutron flux reduction by a 40~cm-thick wall assembled of Boron loaded polyethylene bricks stockpiled in the same room. 
\end{itemize}

In order to estimate a shadowing effect from the each factor we used the GEANT4 model describing the scintillator material, neutron detector box and R1250 PMTs. Exact composition of the borosilicate glass used by Hamamatsu Photonics is unknown \cite{Westerdale2017}, and we used the Pyrex borosilicate glass containing 13\% of B$_2$O$_3$ \cite{BGlass} for modelling of phototubes. In the model, thermal neutrons were generated isotropically in all directions from the surface of semi-sphere a 5~m in diameter. Number of neutrons passing through the scintillator layer positioned alone was used as a reference. We found that the aluminum box reduced number of thermal neutrons capable of reaching the scintillator by 4.8~$\pm$~0.7(stat)\%, while fraction of neutrons blocked by the PMT glass was 3.6~$\pm$~0.7(stat)~\%. Reduction of the thermal neutron flux by the wall made of neutron shielding materials was $\sim$2.0~$\pm$~0.2\%. In total, the thermal neutron flux that passed through the scintillator was suppressed by a factor f~=~0.897$\pm$0.007 compared to the original thermal neutron flux in the underground cavity.

The thermal neutron flux was calculated according to the formula:
\begin{equation}
F[cm^{-2}\cdot sec^{-1}] = \frac{ N_{ev} - R_{\alpha}(s) \cdot t }{\epsilon_{t} \cdot f} \cdot \frac {1}{S \cdot t} 
\label{eq:thermal neutron flux}
\end{equation}
where $N_{ev}$ is the number of events remained after the software box cut demonstrated in Fig.~\ref{fig:neutron_alpha}, $t$ is the time of the measurement in sec, and $S$ is the neutron sensitive area of the detector in cm$^{2}$.
The thermal neutron flux was (6.43$\pm$0.50)$\times$10$^{-6}$ n cm$^{-2}$sec$^{-1}$ that included a $\sim$10~\% contribution from neutrons with energies above 1~eV. The uncertainty in the F value included the statistical uncertainty in $N_{ev}$ (2.1\%) and $R_{\alpha}(s)$ (5.0\%), uncertainty in $\epsilon_{t}$ and f parameters. Within the experimental uncertainties our result is consistent with the thermal neutron flux value (8.26$\pm$0.58)$\times$10$^{-6}$ n cm$^{-2}$sec$^{-1}$ reported by \cite{Nakahata2005}, \cite{Suzuki2009}. A certain difference between the results may originate from technical factors such as deviation of the real $^{241}$Am/$^{9}$Be neutron spectrum from the one reported in \cite{Vijaya1973}, as well as from difference in the Uranium and Thorium concentration in rock and floor in these areas of the Kamioka mine. 

\section{Conclusions}
\label{sec:conclusions}
We successfully developed, constructed and tested a large thermal neutron detector based on 0.32~mm-thick scintillator sheets made of $^6$LiF and ZnS:Ag components. We found that use of a 0.25~mm-thick lamination layer from both sides of the scintillator provides insufficient protection from the airborne radon $\alpha$-activity. To avoid background rate fluctuations and build up of $^{210}$Po in the scintillator we purged out the air containing $^{222}$Rn from the inner detector volume by continuous flow of a pure nitrogen gas. The measured $\alpha$-particle activity in the scintillator before applying the PSD cut was (4.88$\pm$0.06(stat))$\times$10$^{-6}$ $\alpha$ cm$^{-2}$sec$^{-1}$. The neutron detection efficiency was determined by using a low intensity $^{241}$Am/$^{9}$Be fast neutron source with a well known neutron activity measured during deployment at the KamLAND detector. During the neutron detection efficiency measurements the setup consisting of the neutron detector, the neutron source and the water moderator was surrounded by a 10-15~cm-thick neutron shielding that isolated it from surrounding rocks and concrete floor with an uncertain chemical and water content. The PSD discrimination technique was successfully applied to discriminate between neutron signals and the $\alpha$-particle background. Efficiency of the thermal neutron detection was 0.368$\pm$0.018 (0.155$\pm$0.009) before (after) use of the PSD based software selection criteria. The thermal neutron flux at the KamLAND site was (6.43$\pm$0.50)$\times$10$^{-6}$ n cm$^{-2}$sec$^{-1}$ that included a $\sim$10~\% contribution from neutrons with energies above 1~eV. Our result was consistent with the thermal neutron flux value reported by the Super-Kamiokande collaboration in \cite{Nakahata2005}, \cite{Suzuki2009}. Based on the result of the 6~day long measurement using the $^{241}$Am/$^{9}$Be neutron source we concluded that systematic factors such as instability of electronics and PMT gain variations had a very limited effect on detector stability and were insignificant compared with the statistical uncertainty in the number of detected neutrons. Results of all tests demonstrated that our neutron detector can be used to study seasonal variations in the thermal neutron flux at the Kamioka neutrino observatory or similar underground facilities. 




\section*{Acknowledgments}
Development of the thermal neutron detector was supported by the Grant-in-Aid 16K05371 from the Japanese Ministry of Education, 
Culture, Sports, Science and Technology; the World Premier International Research Center Initiative (WPI Initiative), MEXT, Japan. 
We thank Yuri Efremenko from The University of Tennessee for useful advices and comments on the manuscript. We are grateful to Kunio Inoue from The Tohoku University, and Yasuhiro Takemoto from The University of Osaka for their support. 

\end{document}